\documentclass[preprint,double-spaced,showpacs,amsmath,amssymb]{revtex4}

\begin{document}

\title{Dark energy of the Universe
as a field of particles with spin 3}

\author{ B.\,A. Trubnikov}
\email{batrub@nfi.kiae.ru}
\affiliation{Kurchatov Institute, 123182, Moscow, Russia}

\date{\today}

\begin{abstract}
A hypothesis is presented for explanation of the dark matter and
dark energy properties in terms of a new interaction field with
spin 3.

\end{abstract}
 \maketitle

        The quantum physics distinguishes the equations for particles and the equations for
        interaction forces between particles. Elementary (not compound) particles (electrons,
        neutrinoes, quarks) have half-integer spins, S =1/2, whereas interaction fields
        (photons, gravitons and others?) - have the integer spins, S=1, 2, 3,...
        The required quantization for interaction fields' spins leads to cardinal consequences
        for properties of these fields and cooperating particles. In the fields with odd spin
        (magnetic, electric and glueonic fields) the same ``charges'' are repulsive, and different
        ``charges'' are attractive. In contrast to this, in the fields with even spins
        (gravitation field) the same "charges" are attractive and different "charges"
        are repulsive (see[1]). These properties can be named ``the gold rules of Newton-
        Coulomb-Feynman''
        and have been established for the first time in 1935 by E. Vigner, the investigator of the
        quantum nature of spins.

         It is possible to assume on the basis of these ``gold rules'' that the acceleration
         of the Universe expansion (found out in 1998-1999) is caused by the new particles with spin 3.
         Spin 3 causes the repulsion of this particles during there interaction. Let us name them
         conditionally, quints. At the same time, it is reasonable to assume, that there are also
         their anti-particles with an opposite value of the ``charge'' - antiquints. Apparently,
         during the accelerated expansion some part of quints jointed with antiquints, forming
         the pairs, quints + antiquints. They did not annihilated - the absence of the channel
         for annihilation reaction makes impossible for them to turn further.  However, as they
         posses energy and have total even spin $\le6$, these pairs participate in gravitational
         interactions and form  a ``dark matter''. Unlike dark energy (which does not prove itself),
         the dark matter increases observable masses of galaxies (approximately in 10 times)
         forming sometimes ``gravitational lenses''. The ratio between estimated values of dark
         energy and dark matter is equal, approximately.  The theory of ``multi-cluster Bose
         distributions'' discussed below gives just a reasonable explanation for this value.

           It is considered  in this theory, that the set of $N$ particles is broken into
           $K$ clusters of k-type, so the cluster type is defined precisely by number of
           the particles
           which have got into cluster. Therefore, two sums are considered to be preset:
           \begin{equation} N=\sum_k N_k=\sum_k k n_k= const,\,\, and\,\,\,\,    K=\sum_k n_k=
           const,
              \end{equation}
where $n_k$ - number of $k$-type clusters. It is postulated, that
the statistical weight of
 considered set of particles is described by the formula
 \begin{equation} \Omega= \frac{\Gamma(N)}Z,\,\,Z=\prod_k
 Z_k,\,\,Z_k=(k!)^{n_k} \Gamma(N_k).
              \end{equation}
In the simplest case it is possible to take into account only two
types of clusters: singlets with $k=1$, and pairs with $k=2$.
Then the formula (2) becomes simpler,
\begin{equation} \Omega=
\frac{\Gamma(n_1+2n_2)}{\Gamma(n_1)2^{n_2}\Gamma(2n_2)},
              \end{equation}
 where $\Gamma(m) = (m-1!$ - gamma-function of integer arguments.  The requirement of integer
 value for statistical weight $\Omega$ limits the possible choice of integers $n_1$  and $n_2$.
 But, in the
 region of the entropy maximum, $S=\ln \Omega \to max$, these integers can be find by ``manual'' selection
 (``cut-and-try method''). For example, for N=100 the values $n_1 =76,n_2=12$ have been found. This means
 that entropy reaches maximum approximately at $N_1 =n_1 =76$  and $N_2 =24$, at $n_2 =12$. So,
 in this case, the ratio, $N_1/N_2=76/24=3,166$, is rather close to the ratio of ``dark
 energy''  to  ``dark matter'', $  \simeq$ 3.
         We consider this result to be an indirect, but convincing argument for
         our hypothesis of accelerating expansion of the Universe by particles with spin 3.

The author is grateful to E.E. Saperstein for very useful
comments.

{}
\end{document}